\documentclass[aps,prl,twocolumn,showpacs,groupedaddress,amsmath]{revtex4}
\usepackage{graphicx}% Include figure files

\begin{document}

\preprint{PRL/GerigHubler06}

%\preprint{}

\title{Chaos in a one-dimensional compressible flow}


\author{Austin Gerig}
\email{gerig@uiuc.edu}
\author{Alfred H\"{u}bler}
\email{a-hubler@uiuc.edu}

\affiliation{
%\begin{singlespace}
Center for Complex Systems Research, Department of
Physics, University of Illinois at Urbana-Champaign,\\ 1110 West Green Street, Urbana, Illinois 61801
%\end{singlespace}
}


\date{\today}

\begin{abstract}
We study the dynamics of a one-dimensional discrete flow with open boundaries - a series of moving point particles connected by ideal springs. These particles flow towards an inlet at constant velocity, pass into a region where they are free to move according to their nearest neighbor interactions, and then pass an outlet where they travel with a sinusoidally varying velocity.  As the amplitude of the outlet oscillations is increased, we find that the resident time of particles in the chamber follows a bifurcating (Feigenbaum) route to chaos.  This irregular dynamics may be related to the complex behavior of many particle discrete flows or is possibly a low-dimensional analogue of non-stationary flow in continuous systems.
\end{abstract}

%05.45.Ac 	Low-dimensional chaos
%05.45.Pq 	Numerical simulations of chaotic systems
%45.05.+x 	General theory of classical mechanics of discrete systems
%45.50.-j 	Dynamics and kinematics of a particle and a system of particles
%47.10.-g 	General theory in fluid dynamics
%47.52.+j 	Chaos in fluid dynamics

%\keywords{}

\maketitle


%\section{Introduction}
Flows are a frequent topic of research among physicists - fluid flow \cite{Sreenivasan99}, traffic flow \cite{Helbing01}, crowd movement \cite{Helbing00}, and granular flow \cite{Jaeger96} are just a few examples. They are of particular interest because the particles that constitute the flow can exhibit complex behavior at certain flow parameters.  There is a large body of work focused on understanding the cause of this motion and predicting the patterns and structures that these flows produce \cite{Sreenivasan99, Helbing01, Helbing00, Jaeger96, Gollub99}.

Most flows consist of many mutually interacting degrees of freedom and a complete description of the dynamics is often impossible.  It is not surprising then, that theories have historically focused on a statistical description of complex flow \cite{Aref96}. Several studies, however, suggest that we can understand flows at a more fundamental level.  Cellular automata fluid models are an example - for certain types of flows, the scale and particulars of collisions seem unimportant to the overall structure of the flow\cite{Wolfram86,Frisch86}.  The study of bifurcations in Taylor-Couette flow \cite{Brandstater83} suggest that the complicated motion in large scale flows can result from the interplay of a few chaotic degrees of freedom.  Several authors have successfully extended these ideas to the general study of large coherent structures in fluid flow \cite{Holmes96}.  Finally, it has recently been shown that a one-dimensional series of nonlinear oscillators, when driven, can behave quite similar to larger scale turbulent flow \cite{Peyrard02}

We seek to study complex open-boundary flow using a bottom up approach - by studying the simplest possible flows that exhibit complex motion.  The low-dimensional model we present here is unique because particles interact with linear forces and the system has open boundaries.  There are many studies of low-dimensional, \emph{closed}-boundary dissipative systems in the literature.  Examples include driven, damped pendulums, the Lorenz equations, and driven Frenkel-Kontorova models \cite{Sagdeev88, Braun98}.  Like these driven, dissipative systems, the system we present continually gains and depletes energy, but it also allows particles to pass across boundaries into and out of the flow.  Research of complex, yet very low-dimensional, open-boundary systems is scarce in the literature.

We present in this communication a simple one-dimensional flow with open boundaries that exhibits chaotic dynamics.  The system consists of a line of point particles interacting with nearest neighbors according to a linear force law (Hooke's law). These particles travel towards an inlet at constant velocity, pass into a region where they are free to move according to their nearest neighbor interactions, and then pass an outlet where they are driven so that they have a sinusoidally varying velocity.  This outlet driving force continually supplies energy to the system and energy is dissipated when particles exit at any point away from their equilibrium position.  As the amplitude of the outlet oscillations is increased, we find that the resident time of particles between the inlet and outlet follows a bifurcating route to chaos.  We discuss the resulting dynamics of this system and suggest possible implications for larger dimensional flows.

%\section{The System}
\begin{figure}[htb]
\includegraphics[width=3.4in]{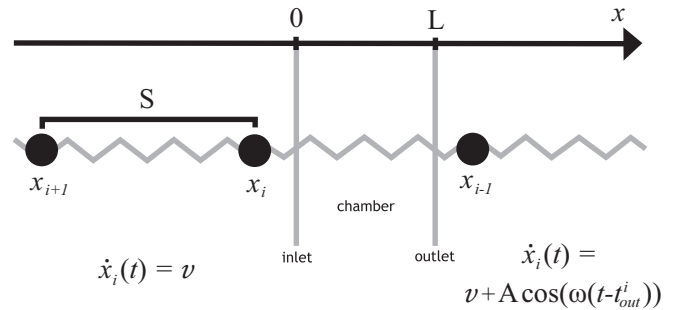}
\caption{\label{fig.system} Diagram of the system.  A series of point particles are connected by ideal springs and initially spaced $S$ apart.  Before reaching $x=0$ (the inlet) and after passing $x=L$ (the outlet), each particle is constrained to the velocities shown.  A particle moves according to its nearest neighbor interactions when between these points.}
\end{figure}
The system consists of a one-dimensional chain of identical point particles spaced a distance $S$ apart and connected by ideal springs.  The particles travel in the $\hat{x}$-direction and their positions are labeled $x_{i}$, where $i$ is the index $i=1,2,\dots,N$.  The number of particles $N$ is chosen large enough so that the flow is sustained throughout our simulations. Particles undergo different dynamics as they pass certain points along the flow. Fig.~\ref{fig.system} is a schematic of the system.

The inlet is located at $x=0$, the outlet at $x=L$, and the region in between is referred to as the chamber.  The time when particle $i$ reaches $x=0$ is labeled $t^{i}_{in}$ and is calculated $t^{i}_{in}=i S/v$.  The time when particle $i$ reaches $x=L$ is labeled $t^{i}_{out}$ and is determined implicitly from the equation $x_{i}(t^{i}_{out})=L$. The velocity of a particle before and after these times is constrained as given in Eqs.~(\ref{eq.vin},\ref{eq.vout}). Particle velocities are not constrained between $t^{i}_{in}$ and $t^{i}_{out}$.
\begin{subequations}
\begin{eqnarray}
\label{eq.vin}
\dot{x}_{i}=v &  & t < t^{i}_{in},\\
\label{eq.vout}
\dot{x}_{i}=v+A \cos(\omega (t-t^i_{out})) & & t\geq t^{i}_{out}.
\end{eqnarray}
\end{subequations}
Here $A$ and $\omega$ are the amplitude and frequency of velocity oscillations after reaching the outlet.

The initial spacing between particles is $S$, a value chosen such that the time between particles entering the chamber is much greater than the average resident time of individual particles within the chamber. This ensures that only one particle is unconstrained at any moment in time.

The following are the complete equations of motion for the particles.
\begin{subequations}
\begin{eqnarray}
x_{i}=v t-i S & & t < t^{i}_{in},\\
m \ddot{x}_{i}+2k x_{i}=k(x_{i-1}+x_{i+1}) & & t^{i}_{in}\leq t < t^{i}_{out},\\
\dot{x}_{i}=v+A \cos(\omega (t-t^i_{out})) & & t\geq t^{i}_{out}.
\end{eqnarray}
\end{subequations}

All particles have the same mass, $m$, and the linear restoring force, $k$, is the same for all springs. The solution for the position of the $i_{th}$ particle for times $t^{i}_{in}\leq t \leq t^{i}_{out}$ is given in Eq.~(\ref{eq.xsol}). It is used to solve $x_{i}(t^{i}_{out})=L$ implicitly for $t^{i}_{out}$ using Newton's method.
\begin{widetext}
\begin{equation}
\label{eq.xsol}
x_{i}(t)=B_1 \cos(\sqrt{2\alpha}t)+B_2 \sin(\sqrt{2\alpha}t) + \frac{A}{\omega(2-\alpha\omega^2)}\sin(\omega(t-t^i_{out}))+vt+(L-S)/2
\end{equation}
\end{widetext}
where,
\begin{subequations}
\begin{eqnarray}
\alpha \; & = & m/k \\
B_1 & = & \frac{(S-L)}{2}-\frac{A}{\omega(2-\alpha\omega^2)}\sin(-\omega t^i_{out}) \\
B_2 & = & -\frac{A}{\sqrt{2\alpha}(2-\alpha\omega^2)}\cos(-\omega t^i_{out})
\end{eqnarray}
\end{subequations}

%\section{Results}
In all of the following results, we use the following parameters: $S=100$, $L=2$, $v=10$, $\alpha=.06$, $\omega=13.2$, $A=0$ to $40$.  These parameter values ensure that, at most, only one particle is located in the chamber at any moment in time.

For each particle we calculate the resident time within the chamber. This quantity is defined,
\begin{equation}
t^{i}_{res} = t^{i}_{out}-t^{i}_{in}.
\end{equation}
$t^i_{in}$ is determined from Eq.~\ref{eq.vin} and $t^{i}_{out}$ is determined by numerically solving Eq.~\ref{eq.xsol} for $x_{i}(t^{i}_{out})=L$.

\begin{figure}[htb]
\includegraphics[width=3.4in]{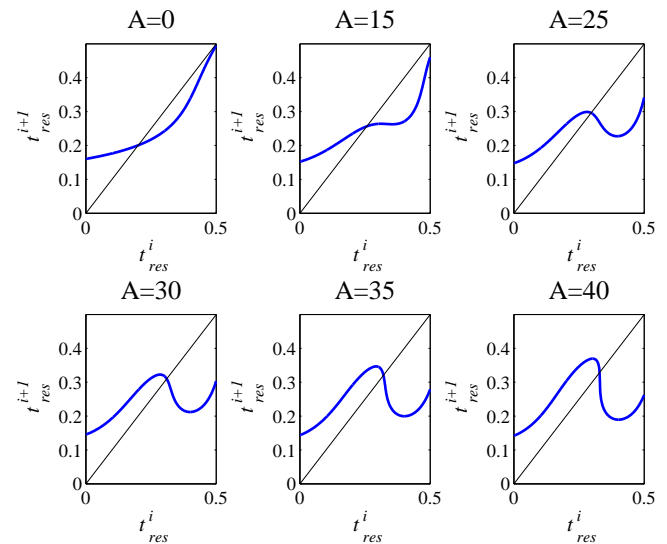}
\caption{\label{fig.returnmap} Return map for $t^{i}_{res}$ for several values of the amplitude, $A$. Parameters for this simulation were: $S=100$, $L=2$, $v=10$, $\alpha=.06$, $\omega=13.2$.}
\end{figure}

Fig.~\ref{fig.returnmap} shows several return maps for the resident time (these map $t^{i}_{res}$ to $t^{i+1}_{res}$).  In general $t^{i+1}_{res}=f(t^{i}_{res},t^{i-1}_{res},...)$, but for this system, 
\begin{equation}
t^{i+1}_{res}=f(t^{i}_{res}), 
\end{equation}
i.e., the return map is one-dimensional.  This results from constraining particle velocities before $x=0$ and after $x=L$ - reducing the system to only one degree of freedom. 

\clearpage

\begin{widetext}
\begin{figure*}[htb]
\includegraphics[width=7in]{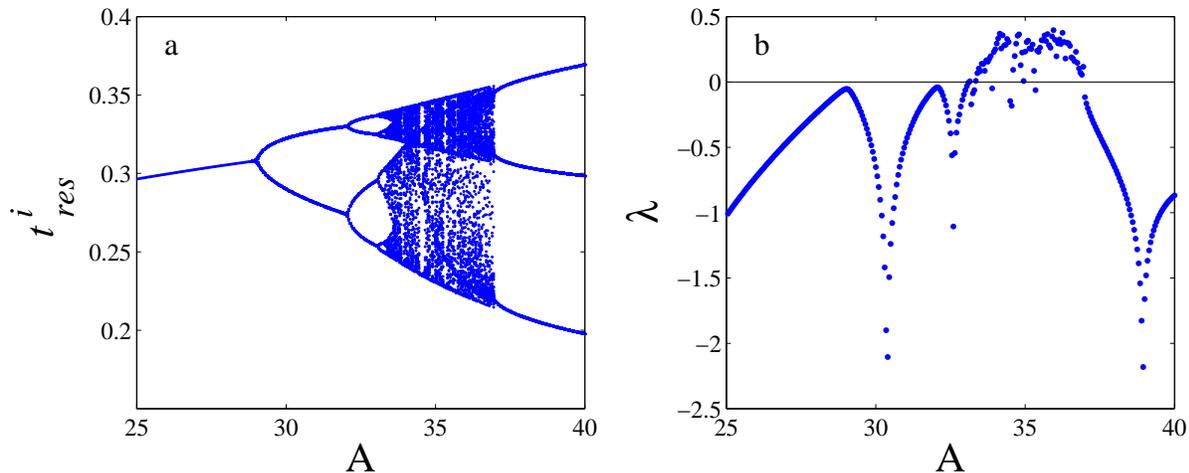}
\caption{\label{fig.bifurcation_lyapunov} a) Bifurcation diagram of $t^{i}_{res}$ plotted for values of $A$ from 25 to 40. b) The Lyapunov exponent, $\lambda$, as a function of the amplitude of outlet oscillations, $A$, also from 25 to 40.  Parameters for this simulation were: $S=100$, $L=2$, $v=10$, $\alpha=.06$, $\omega=13.2$.  The first three bifurcation points are located at $A=29.0$, $A=32.1$, and $A=33.1$; and $\lambda>0$ first at $A=33.4$}
\end{figure*}
\end{widetext}

As $A$ is increased, a peak develops in the return map and the fixed point eventually becomes unstable.  The one-hump map that develops is likely to exhibit the features of many other unimodal return maps - specifically a bifurcation route to chaos \cite{Feigenbaum78}. As Fig.~\ref{fig.bifurcation_lyapunov}a shows, this is indeed the case. The system is iterated onto the attractor and the next 100 values for $t^{i}_{res}$ are plotted for values of the amplitude, $A$, from 25 to 40. The resident time bifurcates several times and eventually becomes chaotic before settling back to a period three dynamics.  The first three bifurcation points are located at $A=29.0$, $A=32.1$, and $A=33.1$.

In Fig.~\ref{fig.bifurcation_lyapunov}b we plot the Lyapunov exponent $\lambda$ for the resident time as a function of the amplitude $A$.  The Lyapunov exponent is a measure of the separation of infinitesimally close trajectories and in this case is calculated numerically from the following equation,
\begin{equation}
\lambda=\lim_{n \rightarrow \infty}\frac{1}{n}
\sum_{i=0}^{n-1}{\log\left|f'(t^{i}_{res})\right|}.
\end{equation}
When $\lambda>0$ trajectories exponentially diverge, which produces chaos when the trajectories remain bounded.  The system becomes chaotic at $A=33.4$ where $\lambda$ first turns positive.

%\section{Discussion}
The system we present above is quite different than other one-dimensional particle models in the literature \cite{Peyrard02, Braun98}.  Instead of using nonlinear interactions between particles, the particles in our system interact with linear forces and constraints are applied abruptly at the boundaries.  This shows that complex motion can arise in a flow at the boundary between simple constrained motions without the need for nonlinear interactions between particles.  Many large scale flows contain regions where the dynamics are tightly constrained to regular motion, with complex motion occuring at the boundaries between these regions.  Simple models such as the one we have presented can provide insight into how this behavior develops.

%\section{Conclusions}
Summarizing, we have presented a fully describable one-dimensional flow of point particles connected by ideal springs.  Particle motion is constrained before reaching an inlet and after passing an outlet, and the system is shown to exhibit chaotic dynamics when particles are driven sinusoidally after crossing the outlet.  The outlet driving force continually adds energy to the system.  No drag force is present, but energy is dissipated when particles exit at any point away from their equilibrium positions.  The model can be reduced to a one-dimensional map that produces chaotic dynamics, showing that chaos can occur in flows at the boundary between simple constrained motion, even when particles in the flow interact with linear forces.

This work was supported by the National Science Foundation Grant No. NSF PHY
01-40179, NSF DMS 03-25939 ITR, and NSF DGE 03-38215.

\end{document}